\documentclass[%
 reprint,
superscriptaddress,
twocolumn,
 amsmath,amssymb,
 aps,
 pre,
]{revtex4-2}

\usepackage{amssymb,amsmath,bm}
\usepackage{enumerate}
\usepackage[utf8]{inputenc}
\usepackage{cancel}
\usepackage{ulem}
\usepackage{color}
\usepackage[colorlinks,linkcolor=blue,urlcolor=blue,citecolor=blue]{hyperref}

\usepackage{float}
\usepackage{xr}
\usepackage[toc]{appendix}
\usepackage{graphicx,epstopdf}
\usepackage{tikz}
\usepackage{comment}

\def\be{\begin{equation}}
\def\ee{\end{equation}}
\def\bea{\begin{eqnarray}}
\def\eea{\end{eqnarray}}

\begin{document}
\title{Resetting optimized competitive first-passage outcomes in non-Markovian systems
}
\author{Suvam Pal}
\email{suvamjoy256@gmail.com}
\affiliation{Physics and Applied Mathematics Unit, Indian Statistical Institute, 203 B.T. Road Kolkata, India}
\author{Rahul Das}
\email{rahuldas@imsc.res.in}
\affiliation{The Institute of Mathematical Sciences, CIT Campus, Taramani, Chennai 600113, India \& 
Homi Bhabha National Institute, Training School Complex, Anushakti Nagar, Mumbai 400094,
India}
\author{Arnab Pal}
\thanks{Corresponding author}
\email{arnabpal@imsc.res.in}
\affiliation{The Institute of Mathematical Sciences, CIT Campus, Taramani, Chennai 600113, India \& 
Homi Bhabha National Institute, Training School Complex, Anushakti Nagar, Mumbai 400094,
India}

\begin{abstract}
  We investigate the role of stochastic resetting in non-Markovian systems, where memory effects arise due to slow relaxation, rugged energy landscapes, disordered environments, and molecular crowding. Using the celebrated continuous-time random walk (CTRW) framework, we analyze first-passage processes with multiple competing outcomes and examine how resetting can selectively enhance desired events. We characterize the efficiency of resetting through conditional mean first-passage times (MFPTs) and demonstrate that its impact is highly sensitive to the underlying waiting-time statistics. Furthermore, we derive an inequality that quantifies how resetting controls fluctuations in conditional first-passage times (FPTs), revealing regimes where variability is significantly suppressed. Our results provide a systematic understanding of how long-term memory influences competitive first-passage outcomes and establish resetting as a powerful control mechanism beyond the conventional Markovian setting.
\end{abstract}

\maketitle

\textbf{\textit{Introduction.}}-- 
In classical Markovian transport processes, such as ordinary Brownian diffusion, the system has no memory of its past and the waiting-time distribution (WTD) is exponential, leading to a linear growth of mean-squared displacement (MSD), $(\langle x^2(t)\rangle \sim t)$. However, many real systems deviate from these assumptions. Interactions with complex or disordered environments often generate non-Markovian dynamics, where transition probabilities depend on the time already spent in a given state. These systems are characterized by heavy-tailed WTDs—such as stretched-exponential or power-law forms—that produce non-linear MSD~\cite{bouchaud1990anomalous} scaling, $(\langle x^2(t)\rangle \sim t^\beta)$ with $(0<\beta<1)$, signaling anomalous~\cite{metzler2019brownian} or subdiffusive behavior~\cite{metzler2000random,metzler2004restaurant}. Such memory-driven transport is widespread in nature, appearing in tracer motion through crowded~\cite{sokolov2012models} polymer networks, structural glasses with broadly distributed energy barriers, the motion of a tagged monomer in a polymer chain, hydrodynamic memory in complex fluids, and even beyond physics—in DNA-binding protein searches, neuronal firing statistics with refractory periods, and financial time series exhibiting volatility correlations. Because such non-Markovian~\cite{guerin2012non} processes dominate the dynamics of many realistic physical, biological, and socio-economic systems, their study is essential for accurate modeling of reaction kinetics, first-passage phenomena, and transport in natural environments.


In non-Markovian transport~\cite{sokolov2000levy,sokolov2002solutions} processes, heavy-tailed waiting-time statistics produce broad, fat-tailed FPT distributions because particles may remain immobilized for exceptionally long intervals due to disorder or energetic variations in the medium. According to the big jump principle~\cite{vezzani2019single} introduced by Barkai and collaborators, a single, unusually long trapping event can dominate the fluctuations and determine the far tail of the FPT distribution, leading to large deviations from standard Markovian behavior. Controlling these rare, extreme pauses is therefore essential for improving completion kinetics in realistic systems. Several approaches have been developed to limit such effects, including tempered or truncated power-law waiting times, which introduce natural temporal cutoffs and restore finite statistical moments \cite{metzler2000random}; removal or modification of deep traps to reduce the dominance of extremely long waiting events in glassy or disordered landscapes \cite{condamin2007first, holl2023controls}; external driving or applied fields, which help particles escape traps more readily \cite{bouchaud1992weak}; and intermittent search strategies, where particles alternate between slow, local exploration and faster relocation modes to avoid being dominated by a single unfavorable trap \cite{benichou2011intermittent}. Together, these methods underscore the importance of managing extreme waiting events to optimize first-passage performance in complex, heterogeneous environments.

Much like other control mechanisms, resetting, where the system is intermittently returned to a prescribed state, has emerged as a successful strategy for expediting first-passage events by preventing rare, exceptionally long excursions that dominate completion times \cite{evans2011diffusion, evans2011diffusionopt, evans2013optimal, pal2015diffusion, reuveni2016optimal, pal2017first, chechkin2018random, masoliver2019anomalous, wang2022restoring, evans2020stochastic, tal2020experimental, gupta2022stochastic,kumar2023universal, pal2024channel, biswas2025target, pal2025optimal,sandev2025shear,liang2025ultraslow,trajanovski2025generalized,pal2019first,pal2020search,pal2019landau,roldan2017path,basu2019symmetric, evans2020stochastic,gupta2016resetting, nagar2016diffusion, bodrova2020continuous, boyer2024power, pal2024random, biswas2025target,biswas2025resetting,biswas2026optimal, besga2020optimal, faisant2021optimal, altshuler2024environmental, vatash2025many, paramanick2024uncovering, goerlich2025taming, kundu2025emulating}. Resetting acts to suppress the influence of extreme fluctuations and narrow the distribution of outcomes without requiring detailed microscopic intervention. In this work, we extend this idea to the CTRWs~\cite{montroll1965random,mendez2022nonstandard,mendez2025occupation} that provide a simple yet powerful framework for modeling non-Markovian transport in complex systems. Unlike classical diffusion with exponential waiting times, CTRWs allow arbitrary WTDs, making them well suited to describe anomalous diffusion in disordered and heterogeneous media. Their flexibility has made them central to modeling memory and temporal heterogeneity across physics, chemistry, and biology \cite{kutner2017continuous}. CTRWs naturally capture subdiffusive motion arising from broadly distributed trapping times, as observed in charge transport in amorphous semiconductors and intracellular diffusion in crowded environments \cite{burov2011time}. Extensions of the framework account for fractional diffusion, weak ergodicity breaking, and the influence of temporal correlations on relaxation dynamics. In porous and geological media, CTRWs successfully describe solute spreading \cite{rajyaguru2025rebuttal}, while in soft matter and living cells, they reproduce intermittent single-particle trajectories and statistics of intracellular motion \cite{munoz2021unsupervised}. By integrating stochastic resetting into this framework, we aim to regulate the long trapping events responsible for broad tails, thereby mitigating variability and optimizing first-passage performance in complex environments.

In particular, our focus is on first-passage processes in which multiple outcomes~\cite{szabo1980first,redner2001guide,condamin2005first, elf2007probing,holcman2014narrow} are possible—some favorable and others undesirable~\cite{pal2025universal} in contrast to the case where the completion is indifferent to the nature of the outcome \cite{mendez2022nonstandard}. Multiple first-passage outcomes naturally emerge in soft-matter and disordered systems, where only a subset corresponds to successful performance \cite{dolgushev2025evidence}. In disordered semiconductors~\cite{scher1975anomalous}, porous geological media~\cite{berkowitz1998theory}, and colloids in quenched optical landscapes~\cite{hanes2012colloids}, particles may either reach a productive exit channel or become trapped for long intervals, generating heavy-tailed first-passage statistics. Since many anomalous systems are governed by broad waiting-time statistics~\cite{berezhkovskii2002channel,satija2020broad}, naturally the competing outcomes can also exhibit large variability dominated by rare, long-lived trapping events. A central question, therefore, is how to systematically bias the flux toward~\cite{berezhkovskii2019exact,dagdug2024diffusion} the desired outcomes while suppressing the undesired ones. Beyond improving the likelihood of favorable outcomes in terms of the mean completion time, we also aim to address how resetting has the potential also to significantly reduce fluctuations in completion times, narrowing the first-passage distribution and enhancing reliability.

\begin{figure}
    \centering
    \includegraphics[width=\linewidth]{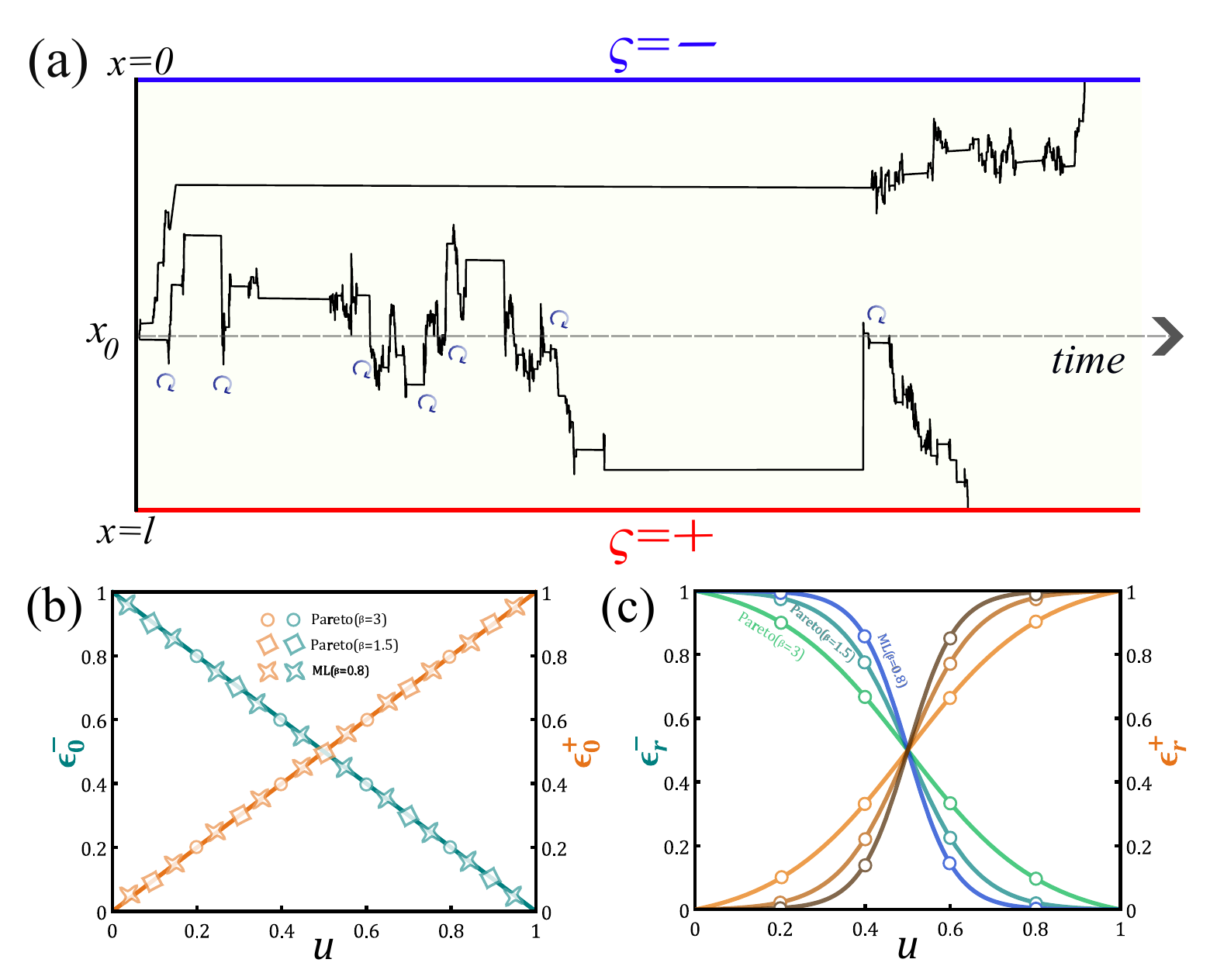}
    \caption{A schematic representation of the CTRW in a one-dimensional arena with two absorbing boundaries with(out) resetting (see panel--(a)). Panel (b) plots the escape probabilities \textit{i.e.,} $\epsilon^-_0=1-u$ and $\epsilon^+_0=u$, which is independent of waiting time statistics. Here we have taken $\tau=0.1$ with different classes of WTD \textit{i.e.,} ML ($\beta=0.8$) and Pareto ($\beta=1.5,3$). However, resetting can reshape these probabilities as shown in panel (c), also breaking the universality.}
    \label{sp_fig}
\end{figure}


\textbf{\textit{Model Set-up details.}} We assume the CTRW to be confined in a one-dimensional arena in the presence of two absorbing boundaries at $x=0$ and $x=l$. The absorbing events through these boundaries are respectively denoted with two mutually exclusive outcomes  \textit{i.e.}, $\varsigma=\{-,+\}$ (see Fig.~\ref{sp_fig}(a)). Initiating from $x_0$, the walker undergoes a series of transitions characterized by random waiting times followed by discrete jumps. This \textit{wait-and-jump} sequence effectively models the continuous interaction between the particle and its environment. In particular, we consider the jump lengths and the waiting times to be independent and identically distributed random variables according to the probability density functions (PDFs) $\psi(x)$ and $\phi(t)$, respectively. Further, we assume the characteristic jump length ($\sigma$) to be small compared to the dimension of the domain \textit{i.e.,} $\sigma \ll l$. While the Markovian case is characterized by the exponential distribution $\phi(t)=\tau^{-1}e^{-t/\tau}$, memory effects emerge as a dominant feature when the WTDs deviate from this exponential form. In what follows, we consider a broad class of $\phi(t)$, characterized by their corresponding Laplace transforms, which mark a strong departure from the Markovianity, namely \cite{metzler2000random}
\begin{equation}\label{waiting-times}
\widetilde{\phi}(s) \approx
\begin{cases}
1 - b_{\beta}\, s^{\beta}, & \text{Class I, } 0 < \beta < 1, \\[3pt]
1 - s \langle \mathcal{T} \rangle_{\phi} + b_{\beta}\, s^{\beta}, & \text{Class II, } 1 < \beta < 2, \\[3pt]
1 - s \langle \mathcal{T} \rangle_{\phi} + \frac{s^{2}}{2} \langle \mathcal{T}^{2} \rangle_{\phi}, & \text{Class III, } \beta > 2.
\end{cases}
\end{equation}
These statistics of random waiting times are associated with the reaction kinetics and often characterize long pauses between successive jumps. This results in diverging mean and second moment (class I), finite mean and diverging second moment (class II) and finite mean and second moment. We seek to see whether intermittent stochastic resetting events can curtail the long trapping intervals (pertinent to class I \& II), and also the class III processes through some $CV$-like criterion. Finally, we derive a general condition that characterizes the impact of resetting on the fluctuations of the random times associated with a specific outcome, extending existing results beyond MFPT optimization.

\begin{figure}
    \centering
    \includegraphics[width=\linewidth]{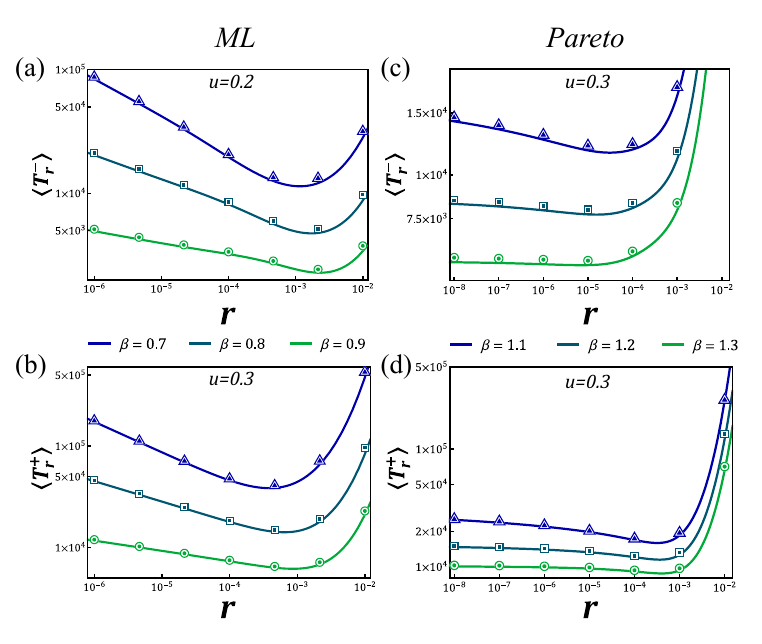}
    \caption{Variation of the conditional MFPTs with resetting rate. Panels (a) and (b) represent the resetting-induced conditional MFPT for WTD, which belongs to \textbf{class I} with different exponents $\beta$. While other parameters are fixed to $\sigma=0.01,~\tau=1$ and domain length $l=1$. The solid lines indicate the theoretical results, whereas the numerical results are indicated with markers. Consequently, following the inequalities in Eqs.~\eqref{app-deriv-cond-ml-minus} and \eqref{app-deriv-cond-ml-plus}, here we find resetting to be useful in optimizing the MFPTs. Similarly, with \textbf{class II} WTD, we observe that intermittent restarts become beneficial in regulating MFPTs as captured in panels (c) and (d), and the corresponding discussions have been made in Eqs.~\eqref{slope-pareto-minus} and \eqref{slope-pareto-plus}. Note that here we consider Mittag-Leffler (ML) kind PDF with exponent $0<\beta<1$ as \textbf{class I} WTD, and the Pareto distribution with exponent $1<\beta<2$ as \textbf{class II} WTD.}
    \label{fig2}
\end{figure}

\textbf{\textit{General framework of resetting-mediated CTRW.}}-- 
To delineate the general formalism, let us consider a stochastic process with a set of multiple outcomes, denoted with $\{\varsigma\}$ in the presence of time-dependent resetting with rate $r(t)$. Initiating from $u$, the probability density flux assigned with a particular outcome $\varsigma$, can be written by incorporating the first renewal formalism~\cite{pal2016diffusion}, which reads
\begin{align}\label{current-renew-main}
    J_r^\varsigma(u,t)=&e^{-R(t)}J_0^\varsigma(u,t)\nonumber\\
    &+\int_0^t d\tau_f~r(\tau_f)~e^{-R(\tau_f)}Q_0(u,\tau_f) J^\varsigma_r(u,t-\tau_f), 
\end{align}
where $R(t)$ is defined as the cumulative resetting rate \textit{i.e.,} $R(t)=\int_0^t r(z)\,dz$ and $Q_0(u,\tau_f)$ denotes the probability that the process survives up to time $\tau_f$, in the absence of resetting. Additionally, $f_{T_0}(u,t)=-dQ_0(u,t)/dt$ is the unconditional FPT density of the underlying process. The first term in the rhs of Eq.~\eqref{current-renew-main} incorporates contribution from resetting-free trajectories, while the integral represents cumulative contribution from multiple trajectories undergoing resetting.
Performing Laplace transformation on the both sides of Eq.~\eqref{current-renew-main}, we arrive at (see Sec.~S1 in Supplemental Material (SM))
\begin{equation}\label{current-renew-main-lap}
    \widetilde{J}_r^\varsigma(u,s) = \dfrac{\widetilde{\mathcal{A}}^\varsigma_0(u,s)}{1-\widetilde{\mathcal{K}}_{Q_0}(u,s)},
\end{equation}
with the following quantities $\widetilde{\mathcal{A}}^\varsigma_0(u,s)=\int_0^\infty e^{-st} e^{-R(t)}J^\varsigma_0(u,t) dt$, and $\widetilde{\mathcal{K}}_{Q_0}(u,s)=\int_0^\infty r(t) e^{-st} e^{-R(t)}Q_0(u,t) dt$.
For the case of exponential resetting, we arrive at
\cite{jain2023fick,pal2024channel,pal2025universal,pal2025optimal}
\begin{equation}\label{current-lap1}
    \widetilde{J}^\varsigma_r(u,s)=\dfrac{\widetilde{J}^\varsigma_0(u,r+s)}{1-r\widetilde{Q}_0(u,r+s)}.
\end{equation}
Furthermore, recalling the relation between the conditional FPT and the corresponding current, one can write \cite{jain2023fick,pal2025universal} $f_{T^\varsigma_{r(0)}}(u,t) = \dfrac{J^\varsigma_{r(0)}(u,t)}{\epsilon^\varsigma_{r(0)}(u)}$, where $\epsilon^\varsigma_{r(0)}(u)=\int_0^\infty J^\varsigma_{r(0)}(u,t)\,dt$ is the exit or splitting probability that starting from $u$, the process exits through a specific boundary in the presence (absence) of resetting. 
Skipping details from SM, we arrive at the following relation connecting  the resetting-mediated conditional FPT density with the underlying FPT densities
\begin{equation}\label{cond-den-res}
    \widetilde{T}^{\varsigma}_r(u,s)=\dfrac{\widetilde{T}^{\varsigma}_0(u,r+s)}{\widetilde{T}^{\varsigma}_0(u,r)}\dfrac{(s+r)\widetilde{T}_0(u,r)}{s+r\widetilde{T}_0(u,r+s)},
\end{equation}
using which we can obtain the conditional MFPT under resetting
\begin{equation}\label{1st-mom-cond-main}
    \langle T^{\varsigma}_r(u)\rangle=\langle T_r(u)\rangle+\dfrac{\partial}{\partial r}\ln \left[\dfrac{\widetilde{T}_0(u,r)}{\widetilde{T}^{\varsigma}_0(u,r)}\right],
\end{equation}
where $\langle T_r(u)\rangle$ represents the unconditional MFPT in the presence of resetting \textit{i.e.,} $\langle T_r(u)\rangle=[1-\widetilde{T}_0(u,r)]/[r\widetilde{T}_0(u,r)]$ \cite{reuveni2016optimal,pal2017first}. Similarly, using Eq.~\eqref{cond-den-res}, one can find the second moment of the resetting-induced conditional FPT density, which reads
{\footnotesize
\begin{align}\label{cond-2nd-moment-main}
    \langle (T^\varsigma_r(u))^2\rangle=\langle T_r(u)^2\rangle&\left[\widetilde{T}_0(u,r)+\dfrac{\langle T^{\varsigma}_r(u)\rangle}{\langle T_r(u)\rangle}(1-\widetilde{T}_0(u,r))\right]\nonumber\\
    &-\left[\dfrac{\dfrac{\partial^2\widetilde{T}_0(u,r)}{\partial r^2}}{\widetilde{T}_0(u,r)}-\dfrac{\dfrac{\partial^2\widetilde{T}^{\varsigma}_0(u,r)}{\partial r^2}}{\widetilde{T}^{\varsigma}_0(u,r)}\right],
\end{align}}
where $\langle T_r(u)^2\rangle=\frac{2(1-\widetilde{T}_0(u,r)+r\partial_r \widetilde{T}_0(u,r))}{r^2\widetilde{T}^2_0(u,r)}$ denotes the second moment of the unconditional FPT density in the presence of resetting.
These expressions, given by Eqs.~\eqref{current-lap1}-\eqref{cond-2nd-moment-main}), require that we first evaluate both the conditional and unconditional FPT quantities for the underlying process. 
For our model system, i.e., the 1D CTRW with two absorbing boundaries \textit{i.e.} $\varsigma=\{-,+\}$, the underlying conditional FPT densities read
\begin{align}
    &\widetilde{T}^-_0(u,s)=\dfrac{1}{1-u}\dfrac{\sinh[(1-u)l\alpha(s)]}{\sinh[l\alpha(s)]},\label{minus-den0}\\
    &\widetilde{T}^+_0(u,s)=\dfrac{1}{u}\dfrac{\sinh[u\alpha(s)]}{\sinh[l\alpha(s)]},\label{plus-den0}
\end{align}
where $u=x_0/l$ and
the corresponding splitting probabilities read $\epsilon^-_0=1-u$ and $\epsilon^+_0=u$, respectively, where $\alpha(s)=\sqrt{2}/\sigma \sqrt{(1/\widetilde{\phi}(s))-1}$. We note that the underlying splitting probabilities are found to be independent of the microscopic details, as illustrated in Fig.~\ref{sp_fig}(b). Irrespective of the time span at a particular location between two successive jumps, the number of trajectories through the left or right exit solely depends on the initial location. Conversely, resetting breaks this symmetry, as can be seen in Fig.~\ref{sp_fig}(c). Using these, we obtain the most general expressions for the conditional MFPT for any WTD
\begin{align}
    \langle T^-_r(u)\rangle &= \langle T_r(u)\rangle + \alpha'[r]\times\dfrac{\mathcal{N}(1-u)}{\mathcal{D}(1-u)},\label{1st-mom-minus-res-1}\\
    \langle T^+_r(u)\rangle &= \langle T_r(u)\rangle + \alpha'[r]\times\dfrac{\mathcal{N}(u)}{\mathcal{D}(u)},\label{1st-mom-plus-res-1}
\end{align}
where $\mathcal{N}(u)=(1-u)l\cosh[(1-u)l\alpha(r)]-ul\coth[ul\alpha(r)]\sinh[(1-u)l\alpha(r)]$ and $\mathcal{D}(u)=\sinh[(1-u)l\alpha(r)]+\sinh[ul\alpha(r)]$.
In what follows, we will discuss the different three classes of WTDs and their intricate role towards the MFPT.

\textbf{\textit{Diverging first and second moments of the WTD}: class I}-- Let us first consider the WTD to be the Mittag-Leffler (ML) kind with $0<\beta<1$, which results in the underlying MFPT being diverging. In the limit $r\rightarrow 0$, the conditional MFPTs read, following \cite{mendez2022nonstandard},
\begin{align}
    \langle T^-_r\rangle &\simeq \dfrac{u(1-u)}{2r}l^2\alpha(r)^2+\mathcal{A}(u,r),\label{app-cond-minus}\\
    \langle T^+_r\rangle &\simeq \dfrac{u(1-u)}{2r}l^2\alpha(r)^2+\mathcal{A}(1-u,r),\label{app-cond-plus}
\end{align}
where $\mathcal{A}(u,r)=\dfrac{l^2}{3}u(2u-1)\alpha(r)\partial_r\alpha(r)$. In particular, WTD in the Laplace domain, which can be expressed as $\widetilde{\phi}(r)=1/[1+(r\tau)^{\beta}]$, which results in $\alpha(r)=\dfrac{\sqrt{2}}{\sigma} (r\tau)^{\beta/2}$. Notably, both conditional MFPTs diverge as $1/r^{1-\beta}$ as $r\to 0$, directly reflecting the heavy-tailed waiting-time statistics. In this context, the efficiency of resetting can be determined by examining the slope of the conditional MFPTs at $r\to 0$,  yielding
\begin{align}
    \left.\partial_r \langle T^{-}_r\rangle\right|_{r\to 0} &\sim -\dfrac{(1-\beta)\tau^\beta}{3r^{2-\beta}}\dfrac{\mathcal{F}(u)l^2}{\sigma^2}<0,\label{app-deriv-cond-ml-minus}\\
    \left.\partial_r \langle T^{+}_r\rangle\right|_{r\to 0} &\sim -\dfrac{(1-\beta)\tau^\beta}{3r^{2-\beta}}\dfrac{\mathcal{F}(1-u)l^2}{\sigma^2}<0,\label{app-deriv-cond-ml-plus}
\end{align}
with $\mathcal{F}(u)=u(3-\beta-u(3-2\beta))>0$. Since the slopes are negative, resets are always beneficial in this case as also can be seen in Fig.~\ref{fig2} (a) \& (b).

\textbf{\textit{Finite first moment and diverging second moment of the WTD}: class II}-- In this case, we consider the WTD to be Pareto distribution with $1<\beta<2$, in the Laplace domain, which can be expressed as $\widetilde{\phi}(r)=\beta~\mathcal{U}(1,1-\beta;r\tau)$. Subsequently, in $r\to 0$, the $\alpha(r)$ reads
\begin{equation}\label{alpha-pareto}
    \alpha(r)\sim \sqrt{r}A_\beta \left[1+B_\beta~r^{\beta-1}\right],
\end{equation}
where $A_\beta=\dfrac{\sqrt{2\tau}}{\sigma\sqrt{(\beta-1)}}$ and $B_\beta=\dfrac{\pi(\beta-1)\tau^{\beta-1}}{2\sin(\pi\beta)\Gamma(1+\beta)}$.
Incorporating Eq.~\eqref{alpha-pareto} in Eqs.~\eqref{minus-den0},\eqref{plus-den0}, conditional MFPTs for the underlying dynamics yield
\begin{equation}
    \langle T^-_{r\to 0}(u)\rangle = \dfrac{(2-u)u}{3(\sigma/l)^2}\langle \mathcal{T}\rangle_\phi,~~~~~\langle T^+_{r\to 0}(u)\rangle = \dfrac{1-u^2}{3(\sigma/l)^2}\langle \mathcal{T}\rangle_\phi,
\end{equation}
with $\langle \mathcal{T}\rangle_\phi=\tau/(\beta-1)$ denoting the first moment of the WTD. Similar to the previous analysis, one can estimate the efficiency of resetting by examining the slope of conditional MFPTs at $r\to 0$. Replacing $\alpha(r)$ with Eq.~\eqref{alpha-pareto} in Eqs.~\eqref{app-cond-minus} and \eqref{app-cond-plus}, one can find the slopes of the conditional MFPTs (see SM, Sec.~S4, Eqs. (S41) and (S42))
\begin{align}
    \left.\partial_r \langle T^{-}_r\rangle\right|_{r\to 0} &\sim \dfrac{(\beta-1)\tau^\beta}{r^{2-\beta}}\dfrac{\Gamma (u)l^2}{6\sigma^2}\dfrac{\pi}{\sin(\pi\beta)\Gamma(1+\beta)}<0,\label{slope-pareto-minus}\\
    \left.\partial_r \langle T^{+}_r\rangle\right|_{r\to 0} &\sim \dfrac{(\beta-1)\tau^\beta}{r^{2-\beta}}\dfrac{\Gamma (1-u)l^2}{6\sigma^2}\dfrac{\pi}{\sin(\pi\beta)\Gamma(1+\beta)}<0,\label{slope-pareto-plus}
\end{align}
with, $\Gamma(u)=u(2-u)>0$ and $\sin(\pi\beta)<0$ for $1<\beta<2$. Thus, reset turns out to be efficient also in this case. This is further verified through Fig.~\ref{fig2} (c) \& (d).

\textbf{\textit{Finite first and second moments of the WTD}: class III}-- The effect of resetting needs to be carefully monitored when the WTD possesses finite first and second moments. In this regime, the efficiency is governed by the following inequality which was derived in \cite{pal2025universal}
\begin{equation}\label{cv0-lam0}
    CV^\varsigma_0(u)>\Lambda^\varsigma_0(u),    
\end{equation}
where $CV^\varsigma_0$ denotes the statistical dispersion of the underlying FPT density for the respective outcome $\varsigma$, and mathematically $\Lambda^\varsigma_0=\left[\frac{\langle T_0\rangle^2}{2\langle T_0^\varsigma\rangle^2}(1+CV_0^2)\right]^{1/2}$. Note that $CV_0$ denotes the statistical dispersion of the underlying FPT irrespective of boundaries. For instance, considering the Pareto WTD with $\beta>2$, this condition leads to the following inequalities 
\begin{align}
    &\mathcal{H}^-(u,\beta)>0,\label{cv0-lam0-left}\\
    &\mathcal{H}^+(1-u,\beta)>0,\label{cv0-lam0-right}
\end{align}
with $\mathcal{H}(u,\beta)=(1-24u+46u^2-19u^3)(\beta-2)+30(1+u)(\sigma/l)^2$, one can find detailed derivation of Eqs.~\eqref{cv0-lam0-left} and \eqref{cv0-lam0-right} in the Sec.~S5 of SM. As described in Fig.~\ref{fig3}(a) the inequality in Eq.~\eqref{cv0-lam0-left} mark the phase domains of efficient versus inefficient resetting. For example, at $u=0.8$ a global minimum in $\langle T_r^-\rangle$ emerges at finite $r$, whereas at $u=0.3$ resetting fails to accelerate the process. An similar behavior can be observed for the escape through the right boundary governed by Eq.~\eqref{cv0-lam0-right}. 

\begin{figure}
    \centering
    \includegraphics[width=\linewidth]{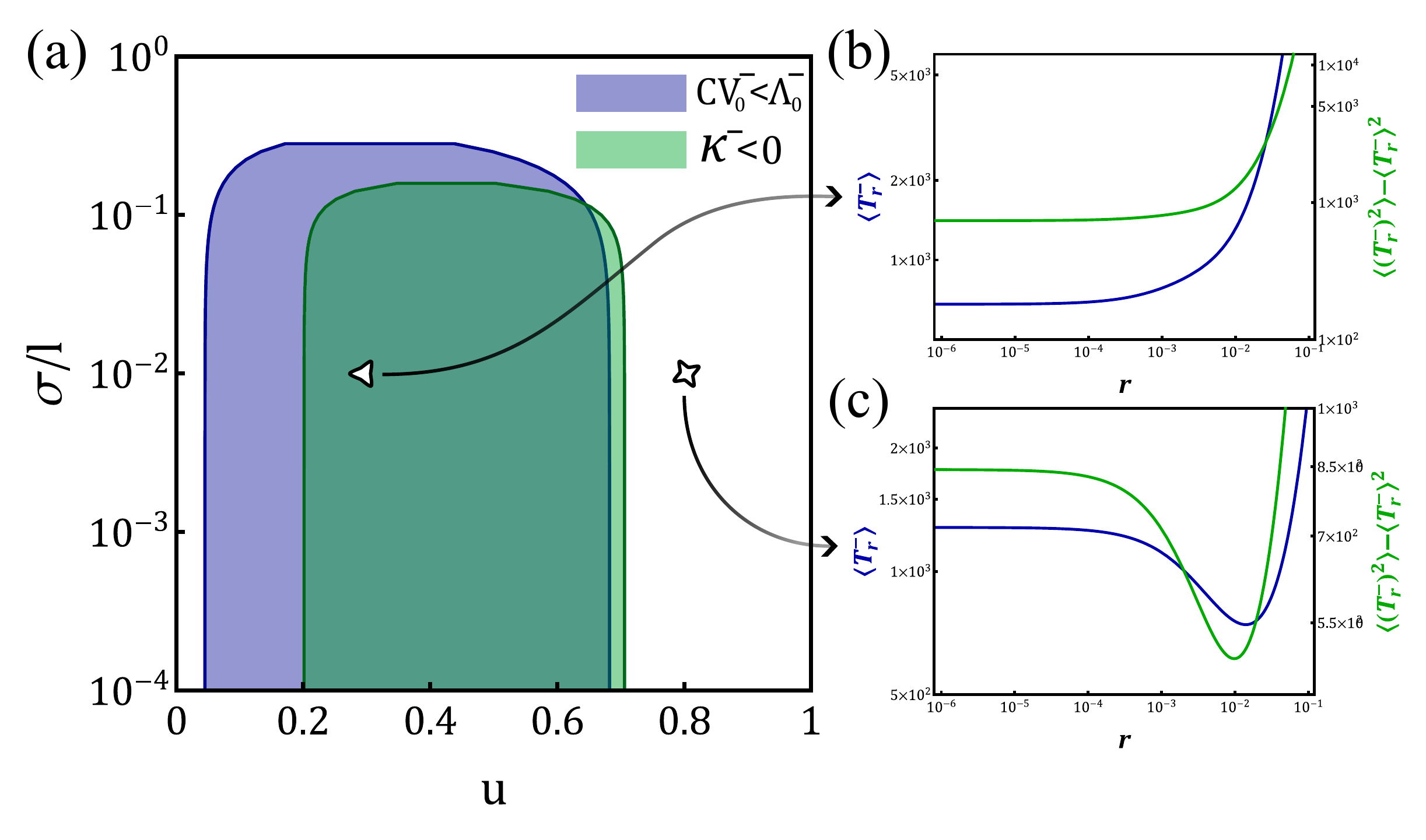}
    \caption{Illustration of the system-parameter domains obtained from the criteria defined in Eqs.~\eqref{cv0-lam0-left} and \eqref{kappa} for realizing the outcome `$\varsigma=-$’, with a Pareto WTD of exponent $\beta=3.5$. Panel (a) shows the regions where $CV_0^- <\Lambda_0^-$ (shaded in violet) and $\kappa^- < 0$ (shaded in green). Panel (b) demonstrates that resetting has a negligible effect on both the conditional MFPT and the exit-time fluctuations through the left boundary for the initial condition $u=0.3$ and domain length $l=1$. In contrast, for the initial condition $u=0.8$, resetting significantly modifies the dynamics and becomes effective in optimizing these metrics, as illustrated in panel (c).}
    \label{fig3}
\end{figure}

\textbf{\textit{Regulation of FPT fluctuations via stochastic resetting}}--
Similar to the mean, intermittent resets can also reduce the fluctuations in the conditional FPTs. By analyzing the fluctuation in $r\to 0$, we obtain the following criterion
\begin{equation}\label{kappa}
	\kappa^{\varsigma}>0,
\end{equation}
with $\kappa^\varsigma=\gamma_1^\varsigma/\widetilde{\gamma}_1-\dfrac{1}{3}\left(\mu_2/\mu_2^\varsigma\right)^{3/2}$, where $\mu^\varsigma_2$ and $\mu^\varsigma_3$ are the second and third central moments of the underlying FPT density $T^\varsigma_0(s)$, respectively. Also, denoting $\widetilde{\gamma}_1$ as standardized third raw moment and the $\gamma_1$ as the central skewness of $\widetilde{T}_0(s)$. For brevity, we provide the derivation of this criterion in Sec.~S6 in the SM. Note that the inequality in Eq.~\eqref{kappa} holds when the underlying matrices are finite i.e., for the class III WTD with $\beta>3$. For classes I and II, resetting always reduces the fluctuations which can be shown by analyzing the slope near $r \to 0$ as was done for the MFPT in above (see Sec. S8 in SM). Eq.~\eqref{kappa} allows one to distinguish the parameter domain, based on the efficiency of intermittent restarts, in terms of the initial conditions $u$ as described in Fig.~\ref{fig3}. For instance, choosing $u=0.3$ (indicated with a triangular marker) in Fig.~\ref{fig3}(b), we demonstrate the behavior of the MFPT and the fluctuation with respect to resetting, linked to the left boundary. Here, resetting is inefficient in controlling the fluctuation in random exit times to exit through the left boundary. Also, resetting mediated MFPT through the left boundary monotonically increases with intermittent restarts. However, with the initial condition $u=0.8$, resetting becomes prominent in optimizing the conditional exit and also controls the fluctuation in this domain -- see Fig.~\ref{fig3}(c). Thus, similar to the conditional $CV^\varsigma$-criterion, the $\kappa^{\varsigma}$ criterion provides a systematic and tunable measure to regulate fluctuations based on the chosen performance metric. In particular, it enables one to identify regimes where fluctuations can be effectively suppressed. This highlights its broader utility as a diagnostic and optimization tool in complex stochastic systems with competing outcomes, especially in the presence of memory and heterogeneity.

\textbf{\textit{Conclusion.}}-- In summary, we have explored how stochastic resetting influences rare-event kinetics in non-Markovian systems characterized by memory, disorder, and anomalous transport. Within the CTRW framework, we have shown that resetting can serve as an effective control mechanism for systems with multiple competing first-passage outcomes, selectively enhancing the likelihood of desired events while suppressing the impact of long trapping episodes. Our analysis of conditional mean first-passage times reveals that the benefits of resetting are strongly dependent on the underlying WTDs, highlighting a nontrivial interplay between memory effects and external control.

A central result of our work is the identification of an inequality governing fluctuations in conditional first-passage times, demonstrating that resetting can significantly reduce variability in regimes dominated by heavy-tailed dynamics. This provides a quantitative framework for understanding how resetting mitigates extreme events and stabilizes system performance. More broadly, our findings extend the applicability of resetting-based optimization strategies beyond Markovian settings, offering new insights into transport, reaction kinetics, and search processes in complex environments.

The framework developed here opens several avenues for future research, including extensions to interacting systems \cite{basu2019symmetric}, spatial resetting protocols \cite{evans2020stochastic,evans2011diffusionopt,roldan2017path}, non-markovian resetting~\cite{gupta2016resetting, nagar2016diffusion, bodrova2020continuous, boyer2024power}, and newly introduced threshold resetting~\cite{biswas2025target,biswas2026optimal}. Given the ubiquity of non-Markovian dynamics in natural and engineered systems, we anticipate that these results will have broad implications for controlling stochastic processes in physics, biology, and beyond.

\bibliography{pal}

\newpage

\begin{titlepage}
    \centering
    {\Large \underline{Supplemental Material} \\[0.5cm]}
    {\Large ``Resetting optimized competitive first-passage outcomes in non-Markovian systems'' \\[0.5cm]}
    {\large Suvam Pal$^{1}$, Rahul Das$^{2}$, and Arnab Pal$^{2}$ \\[0.3cm]}
    {\small 
        $^{1}$ Physics and Applied Mathematics Unit, Indian Statistical Institute, 203 B.T. Road Kolkata, India\\
        $^{2}$ The Institute of Mathematical Sciences, CIT Campus, Taramani, Chennai 600113, India \& 
        Homi Bhabha National Institute, Training School Complex, Anushakti Nagar, Mumbai 400094, India\\
    }
    \vspace{1cm}
\end{titlepage}

\onecolumngrid
\setcounter{page}{1}
\renewcommand{\thepage}{S\arabic{page}}

\setcounter{equation}{0}
\renewcommand{\theequation}{S\arabic{equation}}

\setcounter{figure}{0}
\renewcommand{\thefigure}{S\arabic{figure}}

\setcounter{section}{0}
\renewcommand{\thesection}{S\arabic{section}}

\setcounter{table}{0}
\renewcommand{\thetable}{S\arabic{table}}

\tableofcontents

\section{Derivation of Eqs. (4)-(6) of the main text.}\label{s1}
Consider a resetting mediated first-passage process, starting initially from $u$. Moreover, the dynamics intermittently stop at random time intervals with a time-dependent rate $r(t)$, and re-initiate from the initial state position. Notably, the random intervals of resetting events follow the given distribution $r(t)~exp(-\int_0^t r(z)~dz)$. Since the resetting encodes a repetition of the process at a random interval, using the first renewal formalism, one can write the following \cite{pal2016diffusion}
\begin{equation}\label{surv-renew-general}
    Q_r(u,t)=e^{-R(t)}Q_0(u,t)+\int_0^td\tau_f~r(\tau_f)~e^{-R(\tau_f)}Q_0(u,\tau_f)Q_r(u,t-\tau_f),
\end{equation}
for the survival probability $Q_r(u,t)$ of the resetting-mediated dynamics. Here, the resetting rate $r(t)$ is explicitly dependent on time. Since this is a time-dependent rate process, $e^{-R(t)}$ is the probability of no reset until time $t$. Subsequently, the probability of re-initiating a trajectory between $\tau$ to $\tau_f+d\tau_f$ is $r(\tau_f)e^{-R(\tau_f)}$ for the first time. To quantify the overall behavior of survival probability in the presence of restarts, one has to integrate over time. Since the above relation belongs to the class of Wiener-Hopf integrals, this can be solved in the Laplace domain. By considering the following set of functions
\begin{equation}
    \mathcal{A}_{Q_0}(u,t)=e^{-R(t)}Q_0(u,t),~~\text{and}~~\mathcal{K}_{Q_0}(u,t)=r(t)e^{-R(t)}Q_0(u,t),\nonumber
\end{equation}
with $R(t)=\int_0^t r(z)~dz$, which encodes the following convolution relation
\begin{equation}
    Q_r(u,t)=\mathcal{A}_{Q_0}(u,t)+\left(\mathcal{K}_{Q_0}*Q_r\right)(u,t).
\end{equation}
Note that, $(f*g)(t)=\int_0^t f(\tau_f) g(t-\tau_f)d\tau_f$ denotes the convolution relation over time. Performing Laplace transformation on both sides of the above equation, we get
\begin{equation}\label{surv-lap-general}
    \widetilde{Q}_r(u,s)=\dfrac{\widetilde{\mathcal{A}}_{Q_0}(u,s)}{1-\widetilde{\mathcal{K}}_{Q_0}(u,s)}=\dfrac{\int_0^\infty e^{-st} e^{-R(t)}Q_0(u,t) dt}{1-\int_0^\infty r(t) e^{-st} e^{-R(u,t)}Q_0(u,t) dt}.
\end{equation}
Since we are looking into the resetting-mediated dynamics with a constant rate of resetting $r$, Eq.~\eqref{surv-lap-general} can be further simplified to
\begin{equation}
    \widetilde{Q}_r(u,s)=\dfrac{\widetilde{Q}_0(u,r+s)}{1-r\widetilde{Q}_0(u,r+s)}.
\end{equation}
In terms of the unconditional FPT density $f_{T}(u,t)=-\partial_t Q(u,t)$, the above relation can be further simplified to
\begin{equation}\label{uncond-fpt-lap}
    \widetilde{T}_r(u,s)=\dfrac{(r+s)\widetilde{T}_0(u,r+s)}{s+r\widetilde{T}_0(u,r+s)},
\end{equation}
where $\widetilde{T}_{r(0)}(u,s)$ denotes the unconditional FPT density in the Laplace with(out) resetting. Readily, the first and second moments read
\begin{align}
    \langle T_r(u)\rangle &= \left.-\partial_s \widetilde{T}_r(u,s)\right|_{s\rightarrow 0}=\dfrac{1-\widetilde{T}_0(u,r)}{r\widetilde{T}_0(u,r)},\label{uncond-1st-moment}\\
    \langle T^2_r(u)\rangle &= \left.\partial_s^2 \widetilde{T}_r(u,s)\right|_{s\rightarrow 0}=\dfrac{2}{r^2}\dfrac{1-\widetilde{T}_0(u,r)+r\partial_r\widetilde{T}_0(u,r)}{\widetilde{T}^2_0(u,r)}.\label{uncond-2nd-moment}
\end{align}
Nevertheless, a similar first-renewal equation can be written for the probability density fluxes, which reads
\begin{equation}\label{flux-lap-general}
    J_r^\varsigma (u,t)=e^{-R(t)}J_0^\varsigma (u,t)+\int_0^t d\tau_f~r(\tau_f)~e^{-R(\tau_f)} J^\varsigma_r(u,t-\tau_f)Q_0(u,\tau_f),
\end{equation}
which can be further written by considering the following time-dependent functions
\begin{equation}
    \mathcal{A}^\varsigma(u,t)=e^{-R(t)}J^\varsigma_0(u,t),~~\text{and}~~\mathcal{K}_{Q_0}(u,t)=r(t)e^{-R(t)}Q_0(u,t).\nonumber
\end{equation}
Incorporating these, Eq.~\eqref{flux-lap-general} can be rewritten as
\begin{equation}
    J_r^\varsigma(u,t)=\mathcal{A}^\varsigma(u,t)+\left(\mathcal{K}_{Q_0}*J^\varsigma_r\right)(u,t).
\end{equation}
Performing the Laplace transformation, we get
\begin{equation}
    \widetilde{J}_r^\varsigma(u,s) = \dfrac{\widetilde{\mathcal{A}}^\varsigma(u,s)}{1-\widetilde{\mathcal{K}}_{Q_0}(u,s)}=\dfrac{\int_0^\infty e^{-st} e^{-R(t)}J^\varsigma_0(u,t) dt}{1-\int_0^\infty r(t) e^{-st} e^{-R(u,t)}Q_0(u,t) dt},
\end{equation}
further by taking $r(t)=r$, we arrive
\begin{equation}
    \widetilde{J}_r^\varsigma(u,s) = \dfrac{\int_0^\infty e^{-st} e^{-rt}J^\mp_0(u,t) dt}{1-\int_0^\infty r e^{-st} e^{-rt}Q_0(u,t) dt}=\dfrac{\widetilde{J}^\mp_0(u,r+s)}{1-r\widetilde{Q}_0(u,r+s)}.
\end{equation}
Recalling Eq.~\eqref{surv-lap-general}, in the Laplace domain, Eq.~\eqref{flux-lap-general} reads
\begin{equation}
    \widetilde{J}_r^\varsigma(u,s)=  \widetilde{\mathcal{A}}^\varsigma(u,s)+\widetilde{\mathcal{K}}^\varsigma(u,s)\dfrac{\widetilde{\mathcal{A}}_{Q_0}(u,s)}{1-\widetilde{\mathcal{K}}_{Q_0}(u,s)}.
\end{equation}
Moreover, with a time-independent resetting rate $r(t)=r$, above equation reads
\begin{equation}\label{current-lap}
    \widetilde{J}^\varsigma_r(u,s)=\dfrac{\widetilde{J}^\varsigma_0(u,r+s)}{1-r\widetilde{Q}_0(u,r+s)}.
\end{equation}
Recalling the definition of splitting probabilities \textit{i.e.}, $\epsilon^\varsigma_r(u)=\widetilde{J}^\varsigma_r(u,s\rightarrow 0)$, one can use these in the presence of resetting in the following way
\begin{equation}
    \epsilon^\varsigma_r(u)=\dfrac{\widetilde{J}^\varsigma_0(u,r)}{1-r\widetilde{Q}_0(u,r)}=\dfrac{\widetilde{J}^\varsigma_0(u,r)}{\widetilde{T}_0(u,r)}.
\end{equation}
Substituting $\widetilde{J}^\varsigma_0(u,r)=\epsilon^\varsigma_0(u)\widetilde{T}^\varsigma_0(u,r)$, above relation reads
\begin{equation}\label{splittingr}
    \epsilon^\varsigma_r(u)=\epsilon^\varsigma_0(u)\dfrac{\widetilde{T}^\varsigma_0(u,r)}{\widetilde{T}_0(u,r)}.
\end{equation}
Using Eq.~\eqref{splittingr} upon Eq.~\eqref{current-lap}, the resetting-mediated conditional FPT densities yield
\begin{equation}\label{cond-denr}
    \widetilde{T}^\varsigma_r(u,s)=\dfrac{\widetilde{T}^\varsigma_0(r+s)}{\widetilde{T}^\varsigma_0(r)} \dfrac{(r+s)\widetilde{T}_0(r)}{s+r\widetilde{T}_0(r+s)},
\end{equation}
which is announced in Eq.~\eqref{cond-den-res} in the main text.
Accordingly, the functional form of the first moment reads
\begin{equation}\label{1st-mom-cond}
    \langle T^\varsigma_r(u)\rangle=\langle T_r(u)\rangle+\dfrac{\partial}{\partial r}\ln \left[\dfrac{\widetilde{T}_0(u,r)}{\widetilde{T}^\varsigma_0(u,r)}\right].
\end{equation}
Similarly, the second moment of the conditional FPT in the presence of resetting can be found in terms of the underlying FPT densities, which reads
\begin{equation}\label{cond-2nd-moment-supp}
    \langle (T^\varsigma_r(u))^2\rangle=\langle T_r(u)^2\rangle\left[\widetilde{T}_0(u,r)+\dfrac{\langle T^{\varsigma}_r(u)\rangle}{\langle T_r(u)\rangle}(1-\widetilde{T}_0(u,r))\right]-\left[\dfrac{\dfrac{\partial^2\widetilde{T}_0(u,r)}{\partial r^2}}{\widetilde{T}_0(u,r)}-\dfrac{\dfrac{\partial^2\widetilde{T}^{\varsigma}_0(u,r)}{\partial r^2}}{\widetilde{T}^{\varsigma}_0(u,r)}\right],
\end{equation}
where $\langle T_r^2(u)\rangle = \dfrac{2}{r^2}\dfrac{1-\widetilde{T}_0(u,r)+r\dfrac{\partial \widetilde{T}_0(u,r)}{\partial r}}{\widetilde{T}^2_0(u,r)}$ is the second moment of the resetting-mediated unconditional FPT density \cite{reuveni2016optimal}. Note that, in Sec.~\ref{s3}, we will provide the exact mathematical form of all necessary quantities in terms of the underlying dynamics.

\section{Derivation of splitting probabilities and the FPT densities for the underlying process}\label{s2}
We consider a stochastic process in which a walker undergoes a continuous-time random walk within a finite one-dimensional domain of length $l$, bounded by two absorbing barriers located at $x=0$ and $x=l$. The absorption event in these two barriers can be considered as two mutually exclusive outcomes of a first-passage process, indicated with $\varsigma=\{-,+\}$, \textit{i.e.}, exit pathways through the left and right boundaries, respectively.  We followed Ref.~\cite{mendez2022nonstandard} to derive some of our results below.

Starting from the initial position $x_0$, the walker remains stationary for a random time before making its first jump. After each displacement, it waits for another random time interval before performing the next jump, and this process continues until the walker reaches one of the absorbing boundaries. The jump lengths and waiting times are drawn independently from the respective PDFs $\Psi(x)$ and $\phi(t)$. In particular, we are interested in studying the trade-off between the natural and resetting-mediated absorption, followed by the statistical properties of the WTDs. To study the effect of intermittent restarts on the first-passage process, first, we need to find the conditional FPT  densities without resetting, and furthermore, using those, one can estimate the FPT densities with resetting. 

The propagator $p_0(x,t|x_0)$ that describes the position of the particle at time $t$ given that initially it was located at $x_0$ is given by the famous Montroll-Weiss equation~\cite{montroll1965random}. In the Fourier-Laplace space, this reads
\begin{equation}\label{lap-four-pdf}
    \check{p}_0(k,s|x_0)=\dfrac{e^{ikx_0}[1-\widetilde{\phi}(s)]}{s[1-\widetilde{\phi}(s)\hat{\Psi}(k)]},
\end{equation}
where $\widetilde{[..]}=\mathcal{L}[..]$ and $\hat{[..]}=\mathcal{F}[..]$ denote Laplace and Fourier transformation respectively. Rearranging the above equation, one can write
\begin{equation}\label{lap-four-pdf-2}
    s\check{p}_0(k,s|x_0)-e^{ikx_0}=\widetilde{K}(s)[\hat{\Psi}(k)-1]\check{p}_0(k,s|x_0),
\end{equation}
in Laplace space, the memory kernel is related to the WTD in the following manner
\begin{equation}\label{mem-lap}
    \widetilde{K}(s)=\dfrac{s\widetilde{\phi}(s)}{1-\widetilde{\phi}(s)}.
\end{equation}
Consider the characteristic jump-length $\sigma$ to be small compared to the length of the domain length $l$. This allows the jump statistics to be in the Gaussian limit \textit{i.e.} $\hat{\Psi}(k)\simeq 1-\sigma^2k^2/2$. Finally, inverting Eq.~\eqref{lap-four-pdf-2} in Fourier-Laplace, we obtain the master equation for nonstandard diffusion~\cite{mendez2022nonstandard}
\begin{equation}\label{fp}
    \dfrac{\partial p_0(x,t|x_0)}{\partial t}=\dfrac{\sigma^2}{2}\int_{0}^{t}K(t-t')\dfrac{\partial^2 p_0(x,t'|x_0)}{\partial x^2}~dt',
\end{equation}
with $K(t)=\mathcal{L}^{-1}[\widetilde{K}(s)]$, notably $\mathcal{L}^{-1}[\cdots]$ indicates inverse Laplace transformation. Incorporating the initial condition $p_0(x,t=0|x_0)=\delta(x-x_0)$, in Laplace domain Eq.~\eqref{fp} reads
\begin{equation}\label{fp-lap}
    \dfrac{d^2 \widetilde{p}_0(x,s)}{dx^2}-\dfrac{2s}{\sigma^2\widetilde{K}(s)}\widetilde{p}_0(x,s)=-\dfrac{2\delta (x-x_0)}{\sigma^2\widetilde{K}(s)}.
\end{equation}
Utilizing the boundary conditions $p_0(x=0,t|x_0)=p_0(x=l,t|x_0)=0$, one can find the following solution of the probability density function, which reads
\begin{align}\label{fp-sol}
    \widetilde{p}_0(x,s)&=\dfrac{1}{\sigma}\sqrt{\dfrac{2}{s\widetilde{K}(s)}}\dfrac{1}{\sinh [\alpha(s)l]}\times \begin{cases}
        \sinh[\alpha(s)(l - x_0)] \sinh[\alpha(s)x], & 0 \leq x \leq x_0, \\
        \sinh[\alpha(s)(l - x)] \sinh[\alpha(s)x_0], & x_0 \leq x \leq l,
    \end{cases}
\end{align}
where $\alpha(s)=\dfrac{1}{\sigma}\sqrt{\dfrac{2s}{\widetilde{K}(s)}}=\dfrac{\sqrt{2}}{\sigma}\sqrt{\dfrac{1}{\widetilde{\phi}(s)}-1}$. From Eq.~\eqref{fp}, probability density currents through the right and left boundaries can be written in the following way
\begin{align}
    J^-_0(x_0,t)&=\left.\dfrac{\sigma^2}{2}\int_0^{\infty} K(t-t')\dfrac{\partial p_0(x,t')}{\partial x}\right|_{x\rightarrow 0},\label{left-current}\\
    J^+_0(x_0,t)&=-\left.\dfrac{\sigma^2}{2}\int_0^{\infty} K(t-t')\dfrac{\partial p_0(x,t')}{\partial x}\right|_{x\rightarrow l}.\label{right-current}
\end{align}
Eventually, in Laplace domain above quantities can be written in the following way
\begin{equation}\label{j0-lap}
    \widetilde{J}_0^\mp(x_0,s)=\pm\left.\widetilde{K}(s)\dfrac{\sigma^2}{2}\dfrac{\partial \widetilde{p}_0(x,s)}{\partial x}\right|_{x\rightarrow 0(l)},
\end{equation}
using Eqs.~\eqref{fp-sol}, the probability currents in the Laplace domain read
\begin{align}
    \widetilde{J}^-_{0}(u,s)&=\left.+ \widetilde{K}(s)\dfrac{\sigma^2}{2}\dfrac{\partial \widetilde{p}_0(x,s)}{\partial x}\right|_{x\rightarrow 0}=\dfrac{\sinh[(1-u)l\alpha(s)]}{\sinh[l\alpha(s)]},\label{j0-minus}\\
    \widetilde{J}^+_{0}(u,s)&=\left.- \widetilde{K}(s)\dfrac{\sigma^2}{2}\dfrac{\partial \widetilde{p}_0(x,s)}{\partial x}\right|_{x\rightarrow l}=\dfrac{\sinh[ul\alpha(s)]}{\sinh[l\alpha(s)]}.\label{j0-plus}
\end{align}
Here, we assigned a new variable $u$, which is the scaled initial condition with length of the domain \textit{i.e.}, $u=x_0/l$. In the rest of the article, we will follow this notation. In terms of the probability currents, the splitting probabilities can be written in the following way
\begin{align}\label{epsilon-def}
    \epsilon^-_0(u)&=\widetilde{J}^-_0(u,s\rightarrow 0)=\left.\dfrac{\sinh[(1-u)l\alpha(s)]}{\sinh[l\alpha(s)]}\right|_{s\rightarrow 0},\\
    \epsilon^+_0(u)&=\widetilde{J}^+_0(u,s\rightarrow 0)=\left.\dfrac{\sinh[ul\alpha(s)]}{\sinh[l\alpha(s)]}\right|_{s\rightarrow 0}.
\end{align}
For instance, consider WTD $\widetilde{\phi}(s\to 0)\approx 1-b_\beta s^\beta$ (class-I), where all non-zero moments are diverging. Consequently, $\alpha(s)\approx \dfrac{\sqrt{2}}{\sigma}\sqrt{b_\beta}s^{\beta/2}+\mathcal{O}(s^\beta)$ as $s\to 0$. From Eqs.~\eqref{epsilon-def}, one can expand the $\dfrac{\sinh[(1-u)l\alpha(s)]}{\sinh[l\alpha(s)]}\approx (1-u)+\mathcal{O}(s^\beta)$. Finally, the escape probabilities through the right and left boundaries read
\begin{equation}
    \epsilon^-_0(u)=1-u,~~~~\epsilon^+_0(u)=u.\nonumber
\end{equation}
Which is quite striking irrespective of WTDs and completely universal in this context, as discussed in the main text. Furthermore, the conditional FPT densities in the Laplace domain read
\begin{align}
    &\widetilde{T}^-_0(u,s)=\dfrac{\widetilde{J}^-_{0}(u,s)}{\epsilon^-_0(u)}=\dfrac{1}{1-u}\dfrac{\sinh[(1-u)l\alpha(s)]}{\sinh[l\alpha(s)]},\label{den0-minus}\\
    &\widetilde{T}^+_0(u,s)=\dfrac{\widetilde{J}^+_{0}(u,s)}{\epsilon^+_0(u)}=\dfrac{1}{u}\dfrac{\sinh[ul\alpha(s)]}{\sinh[l\alpha(s)]}.\label{den0-plus}
\end{align}
It is important to note that escape probabilities through the boundaries are independent of the WTD. Also, the unconditional FPT distribution can be written in terms of the weighted sum of the conditional first-passage distributions. Unconditional FPT density yields
\begin{align}
    \widetilde{T}_0(u,s)&=\epsilon^-_0(u)\widetilde{T}^-_0(u,s)+\epsilon^+_0(u)\widetilde{T}^+_0(u,s)=\dfrac{\sinh[(1-u)l\alpha(s)]+\sinh[u\alpha(s)]}{\sinh[l\alpha(s)]}.\label{den0}
\end{align}


\section{Resetting induced FPT densities and respective MFPTs}\label{s3}
In this section, we provide the functional form of the (un)conditional FPT densities with their first moments in the presence of resetting. Using Eq.~\eqref{den0} upon Eq.~\eqref{uncond-1st-moment}, the unconditional FPT reads
\begin{align}\label{ucnond-den-res}
    \widetilde{T}_r(u,s)=\dfrac{r+s}{r+\dfrac{s~\sinh[l\alpha(r+s)]}{\sinh[(1-u)l\alpha(r+s)]+\sinh[ul\alpha(r+s)]}}.
\end{align}
Furthermore, recalling Eq.~\eqref{uncond-1st-moment}, the resetting mediated MFPT irrespective of outcome reads
\begin{align}\label{mfpt-res}
    \langle T_r(u)\rangle=\dfrac{1}{r}\left[\dfrac{\sinh[l\alpha(r)]}{\sinh[(1-u)l\alpha(r)]+\sinh[ul \alpha(r)]}-1\right].
\end{align}
Similarly, one can find the second moment in terms of the underlying unconditional FPT density, which reads
\begin{equation}\label{uncond-den-2nd-supp}
    \langle (T_r(u))^2\rangle = \dfrac{2}{r^2}\dfrac{1-\widetilde{T}_0(u,r)+r\dfrac{\partial \widetilde{T}_0(u,r)}{\partial r}}{\widetilde{T}^2_0(u,r)}.
\end{equation}
Similarly, utilizing Eqs.~\eqref{den0-minus}--\eqref{den0} in Eq.~\eqref{cond-denr}, we find
\begin{align}
    \widetilde{T}^-_r(u,s)=&\dfrac{\sinh[(1-u)l\alpha(r+s)]\sinh[l\alpha(r)]}{\sinh[(1-u)l\alpha(r)]\sinh[l\alpha(r+s)]}\times\dfrac{\sinh[l\alpha(r+s)]}{\sinh[l\alpha(r)]}\nonumber\\
    &\times\dfrac{(s+r)\big[\sinh[ul\alpha(r)]+\sinh[(1-u)l\alpha(r)]\big]}{s\sinh[l\alpha(r+s)]+r\big[\sinh[ul\alpha(r+s)]+\sinh[(1-u)l\alpha(r+s)]\big]},\label{denr-minus}\\
    \widetilde{T}^+_r(u,s)=&\dfrac{\sinh[ul\alpha(r+s)]\sinh[l\alpha(r)]}{\sinh[ul\alpha(r)]\sinh[l\alpha(r+s)]}\times\dfrac{\sinh[l\alpha(r+s)]}{\sinh[l\alpha(r)]}\nonumber\\
    &\times\dfrac{(s+r)\big[\sinh[ul\alpha(r)]+\sinh[(1-u)l\alpha(r)]\big]}{s\sinh[l\alpha(r+s)]+r\big[\sinh[ul\alpha(r+s)]+\sinh[(1-u)l\alpha(r+s)]\big]}.\label{denr-plus}
\end{align}
Readily, from Eq.~\eqref{1st-mom-cond}, we get the conditional MFPTs
\begin{align}
    \langle T^-_r(u)\rangle &= \langle T_r(u)\rangle + \alpha'[r]\times\dfrac{\mathcal{N}(1-u)}{\mathcal{D}(1-u)},\label{1st-mom-minus-res}\\
    \langle T^+_r(u)\rangle &= \langle T_r(u)\rangle + \alpha'[r]\times\dfrac{\mathcal{N}(u)}{\mathcal{D}(u)},\label{1st-mom-plus-res}
\end{align}
where
\begin{align}
    \mathcal{N}(u)=&(1-u)l\cosh[(1-u)l\alpha(r)]-ul\coth[ul\alpha(r)]\sinh[(1-u)l\alpha(r)],\nonumber\\
    \mathcal{D}(u)=&\sinh[(1-u)l\alpha(r)]+\sinh[ul\alpha(r)].\nonumber
\end{align}

\section{Supporting derivation of Eqs.~(17) and (18) of the main text}\label{s4}
In this section, we describe the behavior of the MFPT related to a particular outcome for WTD with a finite first moment but a diverging second moment. In the lower resetting rate \textit{i.e.}, $r\to 0$, recalling Eq.~(9) of the main text
\begin{equation}\label{sm-rto0-minus}
    \langle T^-_{r}\rangle\simeq \underbrace{\dfrac{u(1-u)}{2r}l^2\alpha(r)^2}_{F_1}+\underbrace{\dfrac{u(2u-1)}{3}l^2\alpha(r)\partial_r\alpha(r)}_{F_2}.
\end{equation}
Moreover, Eq. (12) reads $\alpha(r)\sim \sqrt{r}A_\beta[1+B_\beta r^{\beta-1}]$, with necessary constants, described in the main text. From this relation, we can write $\alpha(r)^2\sim r A_\beta^2[1+2B_\beta r^{\beta-1}]$ and
\begin{equation}
    \partial_r \alpha(r)\sim \dfrac{1}{2\sqrt{r}}A_\beta[1+B_\beta r^{\beta-1}]+\sqrt{r}A_\beta[(\beta-1)B_\beta r^{\beta-2}].\nonumber
\end{equation} 
The first term in the rhs \textit{i.e.}, $F_1$ of Eq.~\eqref{sm-rto0-minus} can be further simplified to 
\begin{equation}
    F_1\sim \dfrac{u(1-u)}{2}l^2A_\beta^2[1+2B_\beta r^{\beta-1}].\nonumber
\end{equation}
And the second term can be written as
\begin{align}
    F_2\sim &\dfrac{u(2u-1)}{3}l^2\sqrt{r}A_\beta^2[1+B_\beta r^{\beta-1}]\left[\dfrac{1}{2\sqrt{r}}[1+B_\beta r^{\beta-1}]+\sqrt{r}B_\beta (\beta-1)r^{\beta-2}\right] \sim  \dfrac{u(2u-1)}{6}l^2A_\beta^2 [1+2 B_\beta r^{\beta-1}].\nonumber
\end{align}
Finally Eq.~\eqref{sm-rto0-minus} can be written as
\begin{equation}
    \langle T^-_{r}\rangle \sim \dfrac{u(2-u)}{6}l^2A_\beta^2B_\beta r^{\beta-1}.
\end{equation}
Finally, recalling $A_\beta=\dfrac{\sqrt{2\tau}}{\sigma\sqrt{(\beta-1)}}$ and $B_\beta=\dfrac{\pi(\beta-1)\tau^{\beta-1}}{2\sin(\pi\beta)\Gamma(1+\beta)}$, we get
\begin{equation}\label{sup_slope_minus}
    \partial_r \langle T^{-}_r\rangle \sim \dfrac{(\beta-1)\tau^\beta}{r^{2-\beta}}\dfrac{u(2-u)l^2}{6\sigma^2}\dfrac{\pi}{\sin(\pi\beta)\Gamma(1+\beta)}.
\end{equation}
Similarly, one can examine the slope of the conditional MFPT at $r\to 0$, which reads
\begin{equation}\label{sup_slope_plus}
    \partial_r \langle T^{+}_r\rangle \sim \dfrac{(\beta-1)\tau^\beta}{r^{2-\beta}}\dfrac{(1-u^2)l^2}{6\sigma^2}\dfrac{\pi}{\sin(\pi\beta)\Gamma(1+\beta)}.
\end{equation}
This completes the derivation of Eq.~(14) and (15) in the main text.

\section{Mathematical forms of Eqs.~(20) and (21) in the main text}\label{s5}
Starting from the moment generating function $\widetilde{\phi}(s)$, for a given WTD $\phi(t)$ can be expanded in the following way
\begin{equation}\label{expn-time-dist1}
    \left.\widetilde{\phi}(s)\right|_{s\rightarrow 0}\approx 1-s\langle \mathcal{T}\rangle_\phi+\dfrac{1}{2}s^2\langle \mathcal{T}^2\rangle_\phi+\mathcal{O}(s^3),
\end{equation}
where $\langle \mathcal{T}\rangle_\phi$ and $\langle \mathcal{T}^2\rangle_\phi$ are the first and second moments of the WTD. Recalling the definition of $CV_\phi=\sqrt{\dfrac{\langle \mathcal{T}^2\rangle_\phi-\langle \mathcal{T}\rangle^2_\phi}{\langle \mathcal{T}\rangle^2_\phi}}$, the above relation can rewritten as
\begin{equation}\label{expn-time-dist}
    \left.\widetilde{\phi}(s)\right|_{s\rightarrow 0}\approx 1-s\langle \mathcal{T}\rangle_\phi+\dfrac{1}{2}s^2\langle \mathcal{T}\rangle_\phi^2(1+CV_\phi^2)+\mathcal{O}(s^3),
\end{equation}
Furthermore, using the above result $\alpha(s)$ reads
\begin{equation}\label{alpha-form}
    \alpha^2(s)\simeq \dfrac{2\langle T\rangle_\phi}{\sigma^2}s+\dfrac{\langle T\rangle_\phi^2(1-CV_\phi^2)}{\sigma^2}s^2+\mathcal{O}(s^3).
\end{equation}
Notably, one can find the first and second moments exit times through left boundary for the underlying process. From Eq.~\eqref{den0-minus}, one can derive
\begin{equation}\label{moms-minus}
    \langle T^-_0(u)\rangle = \dfrac{(2-u)u}{3(\sigma/l)^2}\langle \mathcal{T}\rangle_\phi,~~
    \langle (T^-_0(u))^2\rangle = \dfrac{u(2-u)}{45(\sigma/l)^4}\langle \mathcal{T}\rangle_\phi^2\left[4+3u(2-u)+15(CV_\phi^2-1)(\sigma/l)^2\right]
\end{equation}

Similarly, from Eq.~\eqref{den0-plus}, first and second moments of exit times through the right boundary read
\begin{equation}\label{moms-plus}
    \langle T^+_0(u)\rangle = \dfrac{1-u^2}{3(\sigma/l)^2}\langle \mathcal{T}\rangle_\phi,~~\langle (T^+_0(u))^2\rangle = \dfrac{1-u^2}{45(\sigma/l)^4}\langle \mathcal{T}\rangle_\phi^2
    \left[(7-3u^2)+15(CV_\phi^2-1)(\sigma/l)^2\right]
\end{equation}

Also from Eq.~\eqref{den0}, first and second moments of the unconditional FPT read
\begin{equation}\label{moms-uncond}
    \langle T_0(u)\rangle=\dfrac{u(1-u)}{(\sigma/l)^2}\langle \mathcal{T}\rangle_\phi,~~\langle (T_0(u))^2\rangle=\dfrac{u(l-u)}{3(\sigma/l)^4}\langle\mathcal{T}\rangle_\phi^2
    \left[(1+u(1-u))+3(CV_\phi^2-1)(\sigma/l)^2\right]
\end{equation}

For instance, we consider the Pareto PDF as a WTD, which reads
\begin{equation}\label{pareto}
    \phi(t)=\dfrac{\beta}{\tau\left(1+\dfrac{t}{\tau}\right)^\beta},~~~~~\beta>0,
\end{equation}
with finite moments up to order $n$ is $\beta>n$. It has first and second moments
\begin{equation}\label{pareto-moms}
    \langle \mathcal{T}\rangle_\phi=\dfrac{\tau}{\beta-1},~~\langle \mathcal{T}^2\rangle_\phi=\dfrac{2\tau^2}{(\beta-1)(\beta-2)},~~~~~\beta>2
\end{equation}
and the coefficient of variation reads
\begin{equation}\label{pareto-cv}
    CV_\phi=\sqrt{\dfrac{\beta}{\beta-2}}.
\end{equation}

Using the following definitions: $CV^\mp_0(u)=\sqrt{\dfrac{\langle (T^\mp_0(u))^2\rangle-\langle T^\mp_0(u)\rangle^2}{\langle T^\mp_0(u)\rangle^2}}$ and $\Lambda^\mp_0(u)=\sqrt{\dfrac{\langle T_0(u)\rangle^2}{2\langle T^\mp_0(u)\rangle^2}(1+CV_0(u)^2)}$. Utilizing Eqs.\eqref{moms-minus}-\eqref{moms-uncond}, one gets the following
\begin{align}
    &CV^-_0(u)=\left[\dfrac{4-4u+2u^2+15(\sigma/l)^2(CV_\phi^2-1)}{5u(2-u)}\right]^{1/2},\label{minus-cv0}\\
    &CV^+_0(u)=\left[\dfrac{2+2u^2+15(\sigma/l)^2(CV_\phi^2-1)}{5(1-u^2)}\right]^{1/2}.\label{plus-cv0}
\end{align}
Also, one can find the $\Lambda^{\mp}_0$ for the conditional escape times
\begin{align}
    &\Lambda^-_0(u)=\sqrt{\dfrac{3}{2}}\left[\dfrac{(1-u)(1+u-u^2+3(\sigma/l)^2(CV_\phi^2-1)}{u(2-u)^2}\right]^{1/2},\label{minus-base0}\\
    &\Lambda^+_0(u)=\sqrt{\dfrac{3}{2}}\left[\dfrac{u(1+u-u^2+3(\sigma/l)^2(CV_\phi^2-1))}{(1-u)(1+u)^2}\right]^{1/2}.\label{plus-base0}
\end{align}
Using Eq.~\eqref{pareto-cv} in Eqs.~\eqref{minus-cv0}, \eqref{minus-base0}, one can find the domain of initial conditions for given $\sigma$ in an interval of length $l$ in terms of the following inequality
\begin{equation}\label{domain-minus}
    (1-24u+46u^2-19u^3)(\beta-2)+30(1+u)(\sigma/l)^2>0,
\end{equation}
for the trajectories through the left exit. Similarly from Eqs.~\eqref{plus-cv0}, \eqref{plus-base0}, we get
\begin{equation}\label{domain-plus}
    (4-11u-11u^2+19u^3)(\beta-2)+30(2-u)(\sigma/l)^2>0,
\end{equation}
for the right exit pathways.

\section{Derivation of the criterion in Eq.~(22) of the main text}\label{s6}
In this section, we provide the detailed derivation of Eq.~(19) of the main text. Starting with the variation (centralized second moment) of the conditional FPT density, one can write 
\begin{align}
    (\Delta T^\varsigma_{\delta r\to 0})^2=(\Delta T^\varsigma_{0})^2
    +\delta r~\dfrac{1}{3}(\langle (T_0)^3\rangle-6\langle (T^\varsigma_0)\rangle^3+9\langle T^\varsigma_0\rangle \langle (T_0^\varsigma)^2\rangle-3\langle (T^\varsigma_0)^3\rangle)
    +\mathcal{O}((\delta r)^2).
\end{align}
Further, recalling the definitions of third central moment \textit{i.e.}, $\mu^\varsigma_3= \langle (T^\varsigma_0-\langle T^\varsigma_0\rangle)^3\rangle$ and the second central moment $\mu^\varsigma_2 = (\Delta T^\varsigma_0)^2= \langle (T^\varsigma_0-\langle T^\varsigma_0\rangle)^2\rangle$ of the underlying FPT density $T^\varsigma_0(s)$. Also, denoting $\widetilde{\gamma}_1=\langle T_0^3\rangle/(\Delta T_0)^3$ as standardized third raw moment and the $\gamma_1=\mu_3/(\Delta T_0)^3$ as the central skewness of $\widetilde{T}_0(s)$, one can write
\begin{align}
    (\Delta T_{\delta r}^\varsigma)^2=(\Delta T_{0}^\varsigma)^2- \delta r~(\mu_2^\varsigma)^{3/2}\widetilde{\gamma}_1\underbrace{\left[\dfrac{\gamma_1^\varsigma}{\widetilde{\gamma}_1}-\dfrac{1}{3}\left(\dfrac{\Delta T_0}{\Delta T_0^\varsigma}\right)^3\right]}_{\kappa^\varsigma}+\mathcal{O}((\delta r)^2).
\end{align}
Remarkably, the following inequality 
\begin{equation}
    \kappa^\varsigma>0,
\end{equation}
underpins the domain of the initial location, where resetting becomes useful in controlling fluctuation in competitive FPTs, as described in Eq.~\eqref{kappa} in the main text. Additionally, one can analyze the behavior of the variation in FPTs irrespective of outcomes
\begin{equation}
    \langle (T_{\delta r})^2\rangle-\langle T_{\delta r}\rangle^2 \equiv (\Delta T_{\delta r})^2= (\Delta T_{0})^2 + \delta r ~ \Delta T_0^3\left[\dfrac{1}{3}\widetilde{\gamma}_1-\gamma_1\right] + \mathcal{O}(\delta r^2).
\end{equation}
For resetting to control the fluctuation \textit{i.e.,} $\Delta T_{\delta r}<\Delta T_{0}$, one should have $\gamma_1>\dfrac{1}{3}\widetilde{\gamma}_1$.

\section{Analogous description of Eqs. (21) and (22) of the main text}
\begin{figure}[h]
    \centering
    \includegraphics[width=0.5\linewidth]{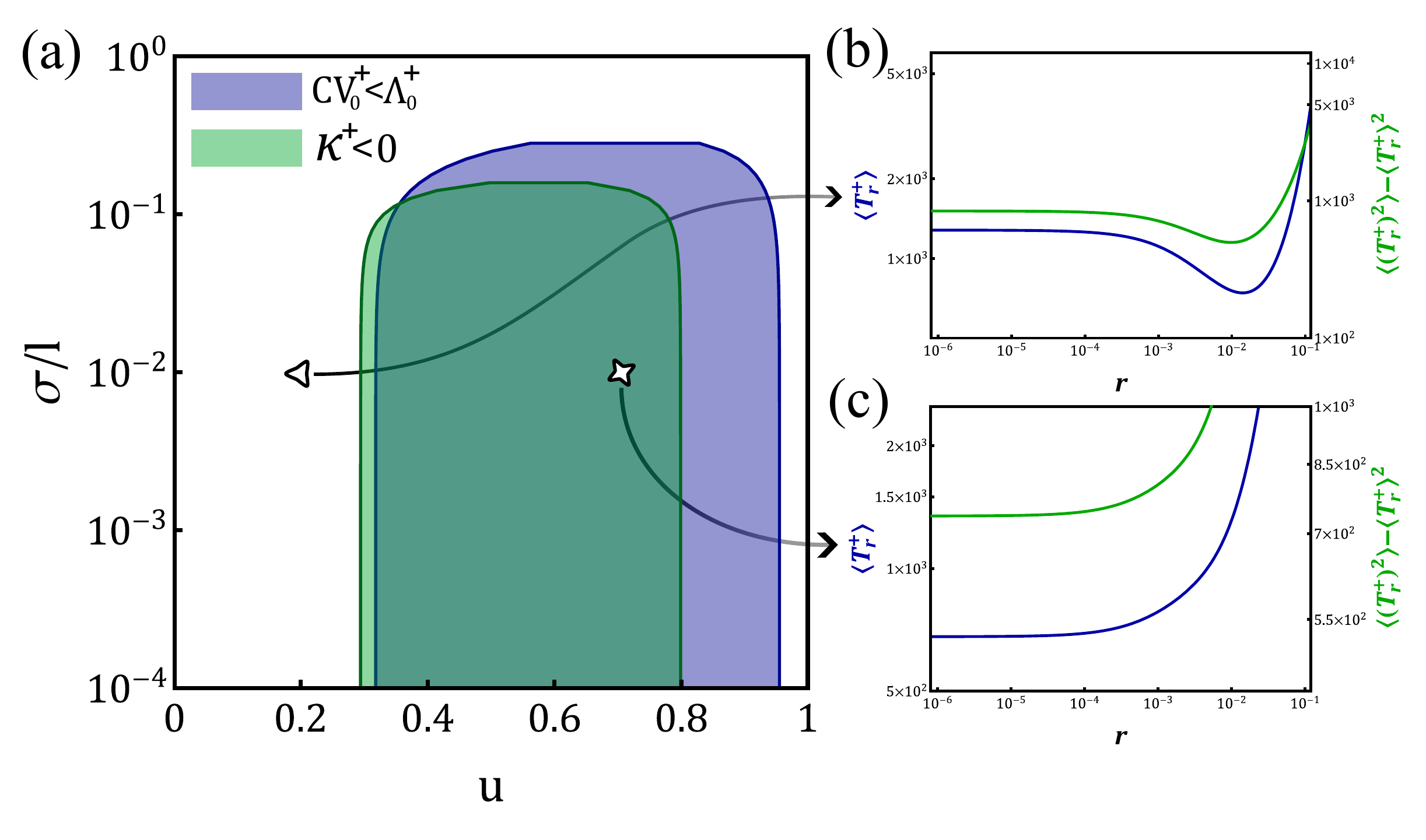}
    \caption{Illustration of the system-parameter domains associated with the realization of the outcome `$\varsigma=+$’, based on the criteria defined in Eqs.~\eqref{cv0-lam0-right} and \eqref{kappa}, for a Pareto WTD with exponent $\beta=3.5$. Panel (a) presents the regions of parameter space where the condition $CV_0^+ < \Lambda_0^+$ is satisfied (shaded in violet), along with the regime $\kappa^+ < 0$ (shaded in green), which together signal parameter ranges where resetting-induced optimization is anticipated. Panel (b) shows that, for the initial condition $u=0.2$ and a fixed domain length $l=1$, stochastic resetting exerts a pronounced influence on both the conditional MFPT and the corresponding exit-time fluctuations through the right boundary. In contrast, for the initial condition $u=0.7$, as illustrated in panel (c), the impact of resetting on the transport properties is significantly reduced, rendering resetting comparatively less effective in this regime.}
    \label{figs1}
\end{figure}
In this section, we provide the parameter domain, obtained from Eq.~\eqref{cv0-lam0-right} \textit{i.e.,} $\mathcal{H}^+<0$ for the conditional MFPT through the right exit. Additionally, we provide the comparison between this inequality with Eq.~\eqref{kappa} \textit{i.e.,} $\kappa^+ < 0$, where resetting is no longer useful in controlling the fluctuation.

\section{Effect of resetting in fluctuations for different WTDs}
In this section, we systematically discuss the behavior of the resetting-mediated fluctuation of the conditional FPTs with different classes of WTDs \textit{i.e.}, $\phi(t)$. We start with recalling the definition of $\alpha(r)=\sqrt{2}/\sigma\sqrt{(1/\widetilde{\phi}(r))-1}$, where $\widetilde{\phi}(r)$ represents the Laplace-transformed WTD. With a given $\alpha(r)$, the centralized second moment of the conditional FPT density can be written as
\begin{align}
    (\Delta T^-_{r\to 0}(u))^2 &\simeq \mathcal{G} (u,r),\label{Del_T_r_minus}\\
    (\Delta T^+_{r\to 0}(u))^2 &\simeq \mathcal{G} (1-u,r),\label{Del_T_r_plus}
\end{align}
where, the analytical form of $\mathcal{G}$ reads
\begin{align}
    \mathcal{G}(u,r) =& \dfrac{1}{3}l^2(1-2u)u(\partial_r\alpha(r))^2+\dfrac{ul^2}{3r}[r(1-2u)\partial_r^2\alpha(r)-6(1-u)\partial_r \alpha(r)]\alpha(r)\nonumber\\
    &+\dfrac{u(1-u)}{r^2}l^2\alpha(r)^2-\dfrac{l^4}{30}u(1-2u)(7-u(19-14u))(\partial_r\alpha(r))^2\alpha(r)^2.\nonumber
\end{align}
To analyze the efficiency of resetting in controlling fluctuation, one needs to probe the slope of the centralized second moment (ref.  Eqs.~\eqref{Del_T_r_minus}, \eqref{Del_T_r_plus}) in the limit $r\to 0$. Importantly, this analysis enables the following relation  
\begin{align}\label{del_G}
    \left.\partial_{r}\mathcal{G}(u,r)\right|_{r\to 0} = & \dfrac{ul^2}{15r^3}\big[15r^2\partial_r\alpha(r)(r(1-2u)\partial^2_r\alpha(r)-2(1-u)\partial_r\alpha(r)\nonumber\\
    &+\alpha(r)^2(l^2r^3(1-2u)(u(19-14u)-7)\partial_r\alpha(r)\partial^2_r\alpha(r)-30(1-u))+r\alpha(r)(60(1-u)\partial_r\alpha(r)\nonumber\\
    &+l^2r^2(1-2u)(u(19-14u)-7)(\partial_r\alpha(r))^3)+5r(r(1-2u)\partial_r^3\alpha(r)-6(1-u)\partial_r^2\alpha(r)))\big].
\end{align}
\textbf{\textit{Class I WTD:}} Let us consider, ML kind WTD with $0<\beta<1$. In the Laplace domain, the WTD can be written as $\widetilde{\phi}(r)=1/(1+(r\tau)^\beta)$, in what follows $\alpha(r)=\frac{\sqrt{2}}{\sigma} (r\tau)^{\beta/2}$. Utilizing this in Eq.~\eqref{del_G}, we get the following
\begin{align}
    \left.\partial_r (\Delta T^-_{r}(u))^2\right|_{r\to 0}&\simeq-\dfrac{l^2\tau^\beta(2-\beta)(1-\beta)}{3r^{3-\beta}\sigma^4}\mathcal{H}(u,\beta),\label{ML_Del_minus}\\
    \left.\partial_r (\Delta T^+_{r}(u))^2\right|_{r\to 0}&\simeq-\dfrac{l^2\tau^\beta(2-\beta)(1-\beta)}{3r^{3-\beta}\sigma^4}\mathcal{H}(1-u,\beta),\label{ML_Del_plus}
\end{align}
where $\mathcal{H}(u,\beta)=u(6-2u(3-\beta)-\beta)$, with $u\in [0,1]$ and $0<\beta<1$ indicate $\mathcal{H}(u,\beta)>0$. Taking this into account, from Eqs.~\eqref{ML_Del_minus} and \eqref{ML_Del_plus}, we find
\begin{equation}
    \left.\partial_r (\Delta T^-_{r}(u))^2\right|_{r\to 0} <0,~~~~~\left.\partial_r (\Delta T^+_{r}(u))^2\right|_{r\to 0}<0.\nonumber
\end{equation}

\textbf{\textit{Class II WTD:}} In particular, we consider Pareto WTD \textit{i.e.}, $\phi(t)=\frac{\alpha}{\tau (1+t/\tau)^{\beta+1}}$ with $1<\beta<2$. In the Laplace domain, this can be expressed as
\begin{equation}
    \widetilde{\phi}(r)=\beta\mathcal{U}(1,1-\beta;r\tau),\label{lap_pareto}
\end{equation}
moreover, in the limit $r\to 0$, we find
\begin{equation}
    \alpha(r)\simeq \dfrac{\sqrt{2}}{\sigma}\dfrac{\sqrt{r\tau}}{\sqrt{\beta-1}}\big[1+\dfrac{\pi(\beta-1)}{2\sin(\pi\beta)\Gamma(1+\beta)}(r\tau)^{\beta-1}\big].\nonumber
\end{equation}
Utilizing this, from Eq.~\eqref{del_G}, one can find the leading terms to be
\begin{align}
    \left.\partial_r (\Delta T^-_{r}(u))^2\right|_{r\to 0}&\simeq \dfrac{\pi l^2\tau^\beta (\beta-1)(2-\beta)}{3r^{3-\beta}\sigma^2\Gamma(1+\beta)\sin(\pi\beta)}\mathcal{H}(u,\beta),\\
    \left.\partial_r (\Delta T^+_{r}(u))^2\right|_{r\to 0}&\simeq\dfrac{\pi l^2\tau^\beta (\beta-1)(2-\beta)}{3r^{3-\beta}\sigma^2\Gamma(1+\beta)\sin(\pi\beta)}\mathcal{H}(1-u,\beta).
\end{align}
Remembering the domain of $\beta$ \textit{i.e.}, $1<\beta<2$ and $u\in (0,1)$ one can find $\mathcal{H}(u,\beta)$ to be positive and $\sin(\pi\beta)<0$. In what follows
\begin{equation}
    \left.\partial_r (\Delta T^-_{r}(u))^2\right|_{r\to 0} <0,~~~~~\left.\partial_r (\Delta T^+_{r}(u))^2\right|_{r\to 0}<0.\nonumber
\end{equation}

\textbf{\textit{Class III WTD:}} Let us consider the Pareto distribution as WTD with $2<\beta<3$, where the first and the second moment of the underlying FPT become finite as described in Eqs.~\eqref{moms-minus} and \eqref{moms-plus}. Recalling Eq.~\eqref{alpha-form}, in the limit $r\to 0$ one can write
\begin{equation}
    \alpha(r)\simeq \dfrac{\sqrt{2}}{\sigma}\dfrac{\sqrt{r\tau}}{\sqrt{\beta-1}}\big[1-\dfrac{1}{2(\beta-2)}r\tau+\dfrac{\pi(\beta-1)}{2\sin(\pi\beta)\Gamma(1+\beta)}(r\tau)^{\beta-1}\big].\nonumber
\end{equation}
Using the above relation, from Eq.~\eqref{del_G}, one gets 
\begin{align}
    \left.\partial_r (\Delta T^-_{r}(u))^2\right|_{r\to 0}&\simeq -\dfrac{\pi l^2\tau^\beta (\beta-1)(\beta-2)}{3r^{3-\beta}\sigma^2\Gamma(1+\beta)\sin(\pi\beta)}\mathcal{H}(u,\beta),\\
    \left.\partial_r (\Delta T^+_{r}(u))^2\right|_{r\to 0}&\simeq -\dfrac{\pi l^2\tau^\beta (\beta-1)(\beta-2)}{3r^{3-\beta}\sigma^2\Gamma(1+\beta)\sin(\pi\beta)}\mathcal{H}(1-u,\beta).
\end{align}
In particular, $0<\sin(\pi\beta)<1$ with $2<\beta<3$, which essentially indicates second centralized moments to be \textit{negative}.

\end{document}